\documentclass{synmet}

\usepackage[dvips]{graphics}

\begin{document}

\def\mbi#1{\mbox{\boldmath$#1$}}
\def\kp{\mbi{k} \cdot \mbi{p}}
\def\eg{E_{\rm g}}
\def\eps{\varepsilon}
\def\la{\langle}
\def\ra{\rangle}
\def\beeq{\begin{equation}}
\def\eneq{\end{equation}}
\def\beeqa{\begin{eqnarray}}
\def\eneqa{\end{eqnarray}}
\def\theequation{\arabic{equation}}
\def\tmptheequation{\arabic{tmpequation}}
\renewcommand{\baselinestretch}{0.85}

\begin{frontmatter}



\title{Excitons in hexagonal nanonetwork materials}


\author{Kikuo Harigaya\thanksref{*}}
\thanks[*]{Corresponding author. Tel: +81-29-861-5151;
Fax: +81-29-861-5375; E-mail: k.harigaya@aist.go.jp}

\address{Nanotechnology Research Institute,
AIST, Tsukuba 305-8568, Japan}

\begin{abstract}
Optical excitations in hexagonal nanonetwork materials, for example,
Boron-Nitride (BN) sheets and nanotubes, are investigated theoretically.  
A permanent electric dipole moment, whose direction is from the B site 
to the N site, is considered along the BN bond.  When the exciton 
hopping integral is restricted to the nearest neighbors, the flat 
band of the exciton appears at the lowest energy. The symmetry of this exciton
band is optically forbidden, indicating that the excitons relaxed
to this band will show quite long lifetime which will cause
luminescence properties.
\end{abstract}

\end{frontmatter}

\renewcommand{\baselinestretch}{0.85}

\section{Introduction}

The hexagonal nanonetwork materials composed of atoms with ionic
characters, for example, Boron-Nitride (BN) sheets and nanotubes [1],
have been investigated intensively.  They are intrinsically
insulators with the energy gap $\Delta$ of about 4 eV as the preceding 
band calculations have indicated [2,3].  The possible photogalvanic 
effects depending on the chiralities of BN nanotubes have been 
proposed by the model calculation [4].

\vspace{3mm}
\begin{center}
\resizebox{!}{5cm}{\includegraphics{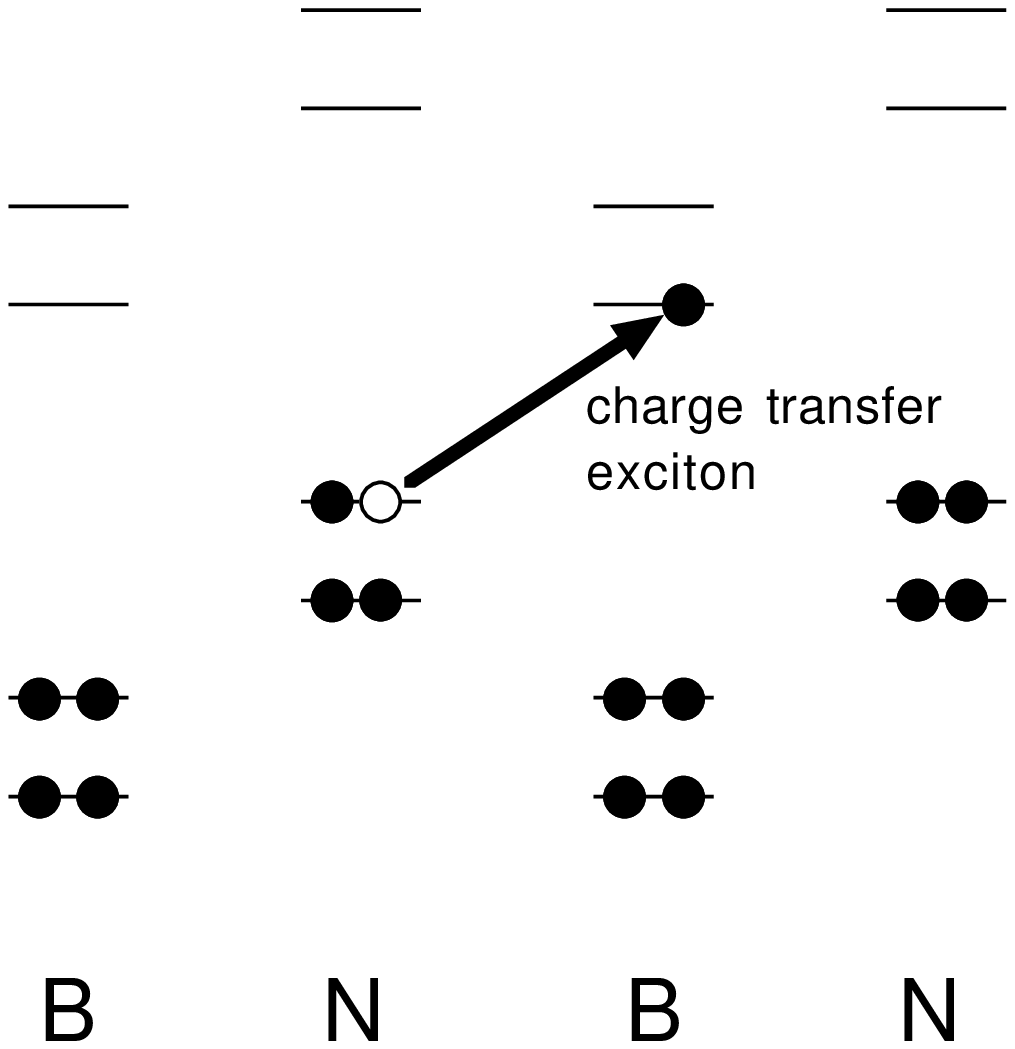}}
\end{center}
\vspace{3mm}

\noindent
{\small Fig. 1. Optical excitations along the BN alternations.}

\vspace{3mm}

In this paper, we investigate optical excitation properties in 
BN systems.   There is a permanent electric dipole moment along 
the BN bond, which will give rise to strong excitonic properties as 
illustrated in Fig. 1.    Low energy optical excitations
are the excitations of the electron-hole pairs between the 
higher occupied states of N and the lower unoccupied states
of B atoms.

\section{Excitons on the Kagom\'{e} lattice}

We consider exciton interactions among nearest neighbor 
dipoles.  In Fig. 2 (a), the B and N atoms are represented by full and
open circles, respectively.  We assume one orbital Hubbard model with
the hopping integral of electrons $t$, the onsite repulsion $U$,
and the energy difference $\Delta$ between the B and N sites.  After
second order perturbations, we obtain the following forms of the 
nearest neighbor interactions: $J_1 = t^2 / (-\Delta + U)$ for the 
case of conserved excited spin (type-1 interaction) and 
$J_2 = t^2/\Delta + t^2/(-\Delta + U)$ for the case that spin of 
the excited electron flips (type-2 interaction).  The condition
$U > \Delta$ means that $J_1$ and $J_2$ are positive.  The interactions 
are present along the thin lines of Fig. 2 (a).  After the extraction 
of the interactions $J_1$ and $J_2$, there remains the two-dimensional 
Kagom\'{e} lattice which is shown in Fig. 2 (b).  Therefore, the 
optical excitation hamiltonian becomes:
\beeqa
H &=& \sum_{\la i,j \ra} \sum_{\sigma = \alpha, \beta} 
J_1 ( |i,\sigma \ra \la j,\sigma | + {\rm h.c.} ) \nonumber \\
&+& \sum_{\la i,j \ra}  
J_2 ( |i,\alpha \ra \la j,\beta | 
+ |i,\beta \ra \la j,\alpha | + {\rm h.c.} ), 
\eneqa
where the indices $i$ and $j$ mean the vertex points 
of the Kagom\'{e} lattice, and the sum is taken over the
nearest neighbor pairs $\la i,j \ra$ and the excited spin $\sigma$.  
The unit cell has three lattice points, 
namely, 1, 2, and 3, as shown in Fig. 2 (b).

\vspace{3mm}
\begin{center}
\resizebox{!}{5cm}{\includegraphics{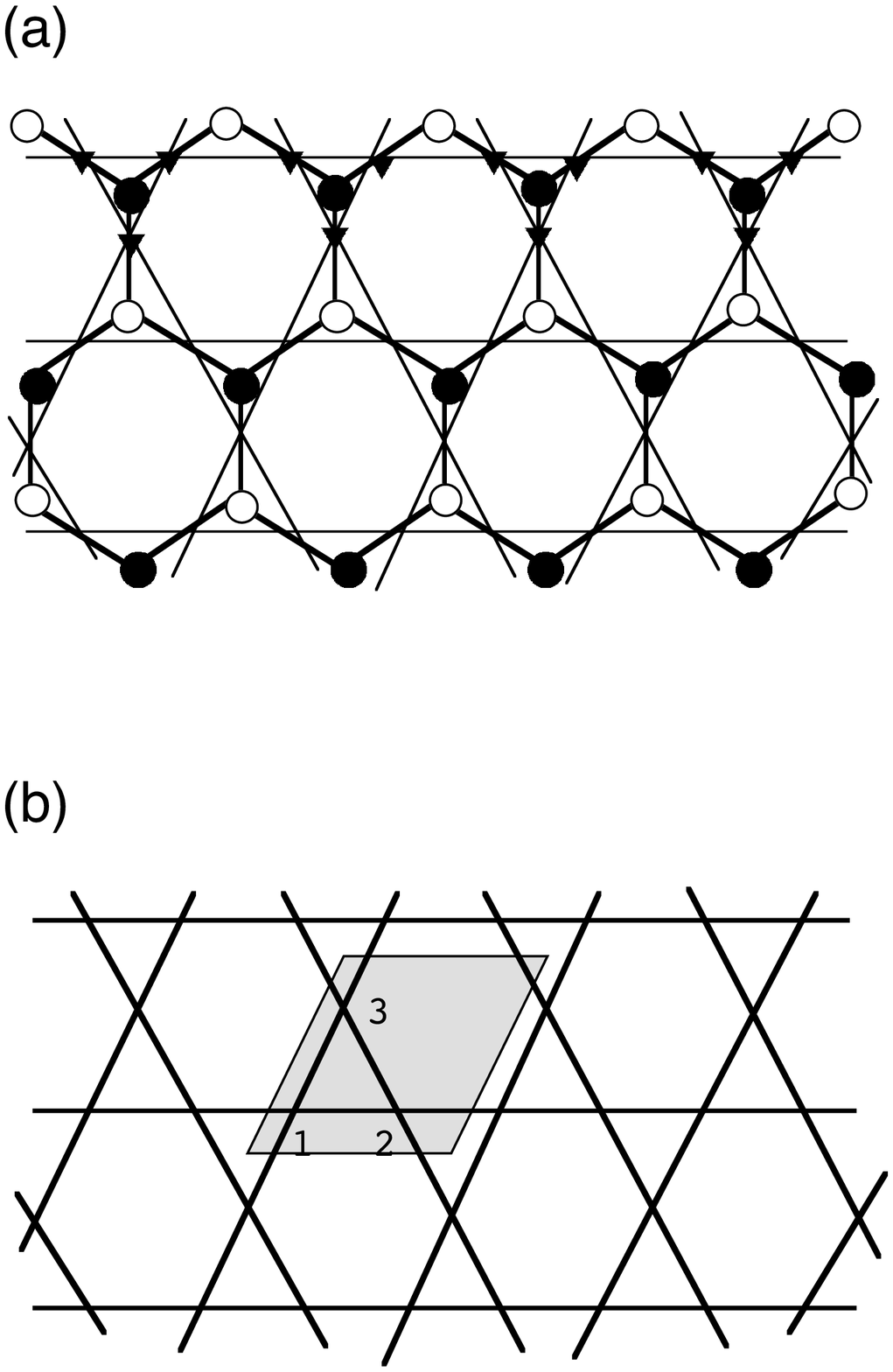}}
\end{center}
\vspace{3mm}

\noindent
{\small Fig. 2. (a) The hexagonal nanonetwork of boron (full circles)
and nitrogen sites (open circles).  Several arrows indicate
the directions of dipole moments. (b) The Kagom\'{e} lattice 
extracted from Fig. (b).  The shaded area is the unit cell.}
\vspace{3mm}

The energy dispersions of the model are given in terms
of wavenumbers $\mbi{k} = (k_x,k_y)$:
\beeq
E =
\left\{ \begin{array}{l}
- 2 (J_1 + J_2), \\
(J_1 + J_2) [1 \pm \sqrt{1 + 4 \cos (k_x b /2)}\\
\overline{\times [\cos (k_x b/2) + \cos (\sqrt{3}k_y b /2)]}],\\
2 (- J_1 + J_2), \\
(J_1 - J_2) [1 \pm \sqrt{1 + 4 \cos (k_x b /2)}\\
\overline{\times [\cos (k_x b/2) + \cos (\sqrt{3}k_y b /2)]}],
\end{array} \right. 
\eneq
where the two dimensional x-y axes are defined as usual
in Fig. 2, and $b = \sqrt{3}a$ is the unit cell length
of the Kagom\'{e} lattice in Fig. 2 (b), and $a$ is the
bond length of Fig. 2 (a).  There appears a dispersionless 
band (triplet state) with the lowest energy $-2 (J_1 + J_2)$.  
There is another dispersionless band (singlet state) at 
the higher energy $2 (- J_1 + J_2)$.  
Such the appearance of the flat band has been discussed
with the possibility of ferromagnetism in the literatures [5].
In the present case, the lowest optical excitation band
becomes flat in the honeycomb BN plane.  When the BN 
plane is rolled up into nanotubes, the flat band is 
dispersionless too.  The flat exciton band will have 
strong optical density originating from the huge density 
of states due to the weak dispersive character.

\vspace{3mm}
\begin{center}
\resizebox{!}{5cm}{\includegraphics{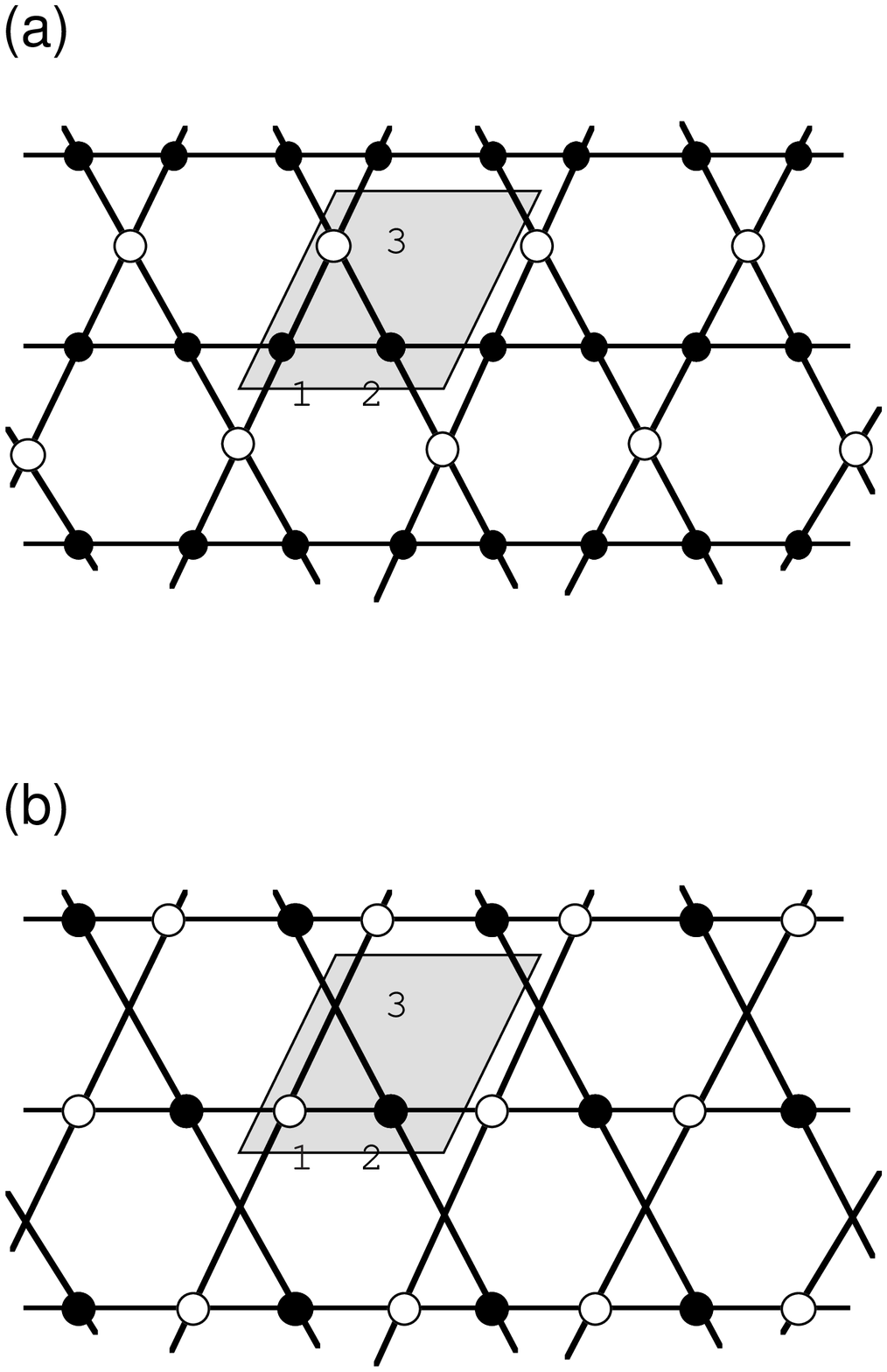}}
\end{center}
\vspace{3mm}

\noindent
{\small Fig. 3.  Symmetries of two wavefunctions at $E=-2(J_1 + J_2)$.
The full and open circles indicate positive
and negative values at the lattice point, respectively.}

\vspace{3mm}

We look at symmetries of the lowest excitons with the
energy $- 2(J_1 + J_2)$ and the wavenumber $\mbi{k} = (0,0)$.
The symmetries of the twofold degenerate solutions are shown in
Fig. 3.  Both wavefunctions have the symmetry {\sl gerade}.  
The transition to the lowest exciton is optically forbidden,
which indicates that excitons relaxed to this lowest exciton band 
will show quite long lifetime which will cause luminescence 
properties.  In addition, the lowest energy excitons will
have huge density of state due to the flatness of the
band.  These properties might result in interesting
optical measurements in hexagonal nanonetwork materials.

\section{Summary}

The flat band of the optically forbidden exciton appears 
at the lowest energy in the optical excitations of BN systems.  
The excitons relaxed to this band might show quite long lifetime 
which will cause luminescence properties.


\begin{thebibliography}{9}

\bibitem{1} D. Golberg, Y. Bando, K. Kurashima, and T. Sato, 
Solid State Commun. 116 (2000) 1.
\bibitem{2} A. Rubio, J. L. Corkill, and M. L. Cohen,
Phys. Rev. B 49 (1994) 5081.
\bibitem{3} X. Blase, A. Rubio, S. G. Louie, and M. L. Cohen,
Europhys. Lett. 28 (1994) 335.
\bibitem{4} P. Kr\'{a}l, E. J. Mele, and D. Tom\'{a}nek,
Phys. Rev. Lett. 85 (2000) 1512.
\bibitem{5} A. Mielke, J. Phys. A 24 (1991) 3311; 
{\sl ibid.} 25 (1992) 4335.
\end{thebibliography}
\end{document}